\documentclass{PoS}

\usepackage{amsmath}
\usepackage{amsfonts}
\usepackage{amssymb}
\usepackage{graphicx}
\usepackage{fontenc}
\usepackage{times}
\usepackage{mathptmx}

\newcommand{\Tr}{\mathrm{Tr}}

\newcommand{\point}{\; .}
\newcommand{\comma}{\; ,}

\newcommand{\LambdaR}{\Lambda_{\rm R}}
\newcommand{\LambdaUV}{\Lambda_{\rm UV}}
\newcommand{\SC}{\mathcal S}
\newcommand{\real}{{\rm Re}\hskip 1pt}

\title{Scalar mass corrections from compact extra dimensions on the lattice}

\ShortTitle{Scalar mass corrections from compact extra dimensions on the lattice}

\author{Luigi Del~Debbio\\
  SUPA, School of Physics and Astronomy, University of Edinburgh\\
  Edinburgh EH9 3JZ, UK\\
  E-mail: \email{luigi.del.debbio@ed.ac.uk}
}
\author{\speaker{Enrico Rinaldi}
  \\
  SUPA, School of Physics and Astronomy, University of Edinburgh\\
  Edinburgh EH9 3JZ, UK\\
  E-mail: \email{e.rinaldi@sms.ed.ac.uk} 
}

\abstract{We explore the phase diagram of the SU($2$) Yang--Mills theory in $5$
dimensions by numerical simulations. The lattice system shows a
dimensionally--reduced phase where the extra dimension is small
compared to the four-dimensional correlation length. In the
low--energy regime of this phase, the system behaves like a
four--dimensional gauge theory coupled to an adjoint scalar field.\\
By tuning the bare parameters of the lattice model, we identify lines
of constant physics, and analyse the behaviour of the non--perturbative scalar
mass as a function of the compactification and the cut--off scales.\\
The perturbative prediction that the effective theory contains a light
particle with a mass that is independent of the cut--off is tested
against non--pertubative results. 
}

\FullConference{ The XXIX International Symposium on Lattice Field Theory - Lattice 2011\\
July 10-16, 2011\\
Squaw Valley, Lake Tahoe, California}

\begin{document}
\section{Introduction}
\label{sect:introduction}

Five--dimensional Yang--Mills theories with an extra dimension
compactified on a circle predict the existence of a scalar low--energy
mode, whose mass renormalisation is suppressed by the remnant of the
higher--dimensional gauge symmetry. It is well known that quantum
corrections yield divergences in the
mass of a scalar particles: the scalar mass receives contributions
proportional to the 
square of the ultra--violet (UV) cut--off. However, the mass of the
scalar particle coming from the compactification of a
higher--dimensional gauge field remains finite, as suggested by
one--loop and two--loop calculations
~\cite{perturbativecalculations}.\\
The five--dimensional gauge theory is perturbatively
non--renormalisable and it can be defined as a 
regulated theory with an ultra--violet cut--off $\LambdaUV$.
In this theory, it is rather surprising to find quantum corrections to the
scalar mass independent of $\LambdaUV$ and taking the form
\begin{equation}
  \label{eq:one-loop-mass}
  m_5^2 \; \equiv \; \delta m^2 \; = \; \frac{9 g_5^2 N_c}{32 \pi^3
    R^3} \zeta (3)
  \comma
\end{equation}
where $\zeta$ is the Riemann Zeta--function, $N_c$ is the numer of
colours, $R$ is the radius of the extra dimension and $g_5^2$ is the
coupling constant of the five--dimensional Yang--Mills
theory. However, Eq.~\eqref{eq:one-loop-mass} is valid
only in the regime where there is a scale separation between the
compactification scale $\LambdaR \sim R^{-1}$ and the cutoff
$\LambdaUV$, because
in this case the details of the regularisation can be neglected.\\
The aforementioned result makes the compactification mechanism a
very interesting and promising scenario to protect the mass of scalar
particles from cut--off effects. This proceeding is a summary of our
recent work~\cite{mypaper} where we studied a simple extra--dimensional
model (cfr. Sec.~\ref{sec:lattice-setup}) regularised on the lattice
with periodic boundary conditions.
Using numerical simulations in the region of phase space where there is a
hierarchy of scales $\LambdaUV \gg \LambdaR$, we are able to study the
parametric dependence of the non--perturbative scalar mass $m_5$ on
the cut--off $\LambdaUV$ and on the compactification scale
$\LambdaR$. This allows us to clarify the status of
Eq.~\eqref{eq:one-loop-mass} in the non--perturbative regime.\\

\section{Scales separation in the lattice model}
\label{sec:lattice-setup}

In recent years, there have been a number of studies on the simplest
of the extra--dimensional theories on the lattice, namely a SU($2$)
pure gauge theory on a five--dimensional torus with anisotropic
couplings~\cite{ejiri,philippe,antonio}.\\
We discretise the pure gauge action using the following
anisotropic lattice Wilson action:
\begin{equation}
  \label{eq:aniso-lattice-action-E}
  \SC_W \; = \; \beta_4 \sum_{x;1\leq \mu <
    \nu \leq 4} \left[ 1 - \frac{1}{2} \real \Tr P_{\mu\nu}(x)
  \right] +
  \beta_5 \sum_{x;1\leq \mu \leq 4}
  \left[ 1 - \frac{1}{2} \real \Tr P_{\mu 5}(x)
  \right] 
  \comma
\end{equation}
where the two lattice bare coupling constants $\beta_4$ and $\beta_5$ can be
tuned independently. This action describes a lattice system with two
independent lattice spacings $a_4$ and $a_5$, corresponding
respectively to the lattice spacing in the 
four--dimensional subspace, and in the extra fifth direction; the
bigger of the two defines the inverse of the cut--off $\LambdaUV$.\\
The anisotropy of the couplings $\gamma = \sqrt{\frac{\beta_5}{\beta_4}}$ is
related to the ratio of the lattice spacings
$\xi = a_4/a_5$. At tree--level $\gamma=\xi$, but quantum corrections
make $\xi$ deviate from this value. The relation between $\xi$ and
$\gamma$ for this action has already been studied in bare parameter
space and can be found interpolating the data of Ref.~\cite{ejiri}. In
the following we restrict ourselves to study $\xi \geq 1$ and the
cut--off is given by $\LambdaUV \sim a_4^{-1}$.\\
Two more parameters in the lattice model can be
adjusted in order to realise the desired separation of scales; they
are $N_4$, the number of lattice sites in the usual four directions,
and $N_5$, the number of lattice sites in the extra
dimension. Together with the corresponding lattice spacings, they
determine the \emph{physical} size of the system: $L_4 = a_4 N_4$ in
four dimensions and $L_5 = 2 \pi R = a_5 N_5$ in the fifth
dimension.\\
Using the model described above, we would like to find a region of its
parameters space where we observe the following:
\begin{itemize}
\item A separation between the compactification scale and the cut--off
  \begin{equation}
    \label{eq:separation-cutoff-radius}
    \LambdaUV \; \gg \; \LambdaR
    \point
  \end{equation}
  This translates into the following relation for the lattice model
  parameters
  \begin{equation}
    \label{eq:separation-lattice}
    \frac{a_5 N_5}{a_4} \; =
    \; \frac{N_5}{\xi} \; \gg \; 1
    \comma
  \end{equation}
  and allows us to rely on the results of the five--dimensional theory
  when describing the low--energy physics.
\item A separation between the four--dimensional physics and the cut--off
  \begin{equation}
    \label{eq:separation-cutoff}
    \sqrt{\sigma} \; \ll \; \LambdaUV \; {\rm ;} \qquad m_5 \; \ll \;
    \LambdaUV
    \point
  \end{equation}
  The above relations translate into
  \begin{equation}
    \label{eq:separation-cutoff-lattice}
    a_4\sqrt{\sigma} \; \ll \; 1\; {\rm ;} \qquad a_4m_5 \; \ll \; 1
    \point
  \end{equation}
  If the above relations are true, we expect the long distance physics
  to be independent of the regularisation details of the theory.  
\item A separation between the four--dimensional physics and the
  compactification scale
  \begin{equation}
    \label{eq:separation-radius}
    \sqrt{\sigma} \; \ll \; \LambdaR \; {\rm ;} \qquad  m_5 \; \ll \;
    \LambdaR
    \point
  \end{equation}
  In terms of the lattice model we have
  \begin{equation}
    \label{eq:separation-radius-lattice}
    \frac{N_5}{\xi} a_4\sqrt{\sigma} \; \ll \; 1 \; {\rm ;} \qquad
    \frac{N_5}{\xi}a_4m_5 \; \ll \; 1
    \point
  \end{equation}
  If this holds, higher Kaluza--Klein modes do not enter the relevant
  dynamics for the low--energy physics.
\item A scalar mass $m_5$ in \emph{physical} units independent of the cut--off
  \begin{equation}
    \label{eq:light-scalar}
    \frac{m_5^2}{\sigma} \; \propto \; \LambdaR^2
    \comma
  \end{equation}
  as expected from Eq.~\eqref{eq:one-loop-mass}.
\end{itemize}

\section{Phase diagram and results from lattice simulations}
\label{sec:results}

\begin{figure}[b]
 \centering
  \includegraphics[width=0.70\textwidth]{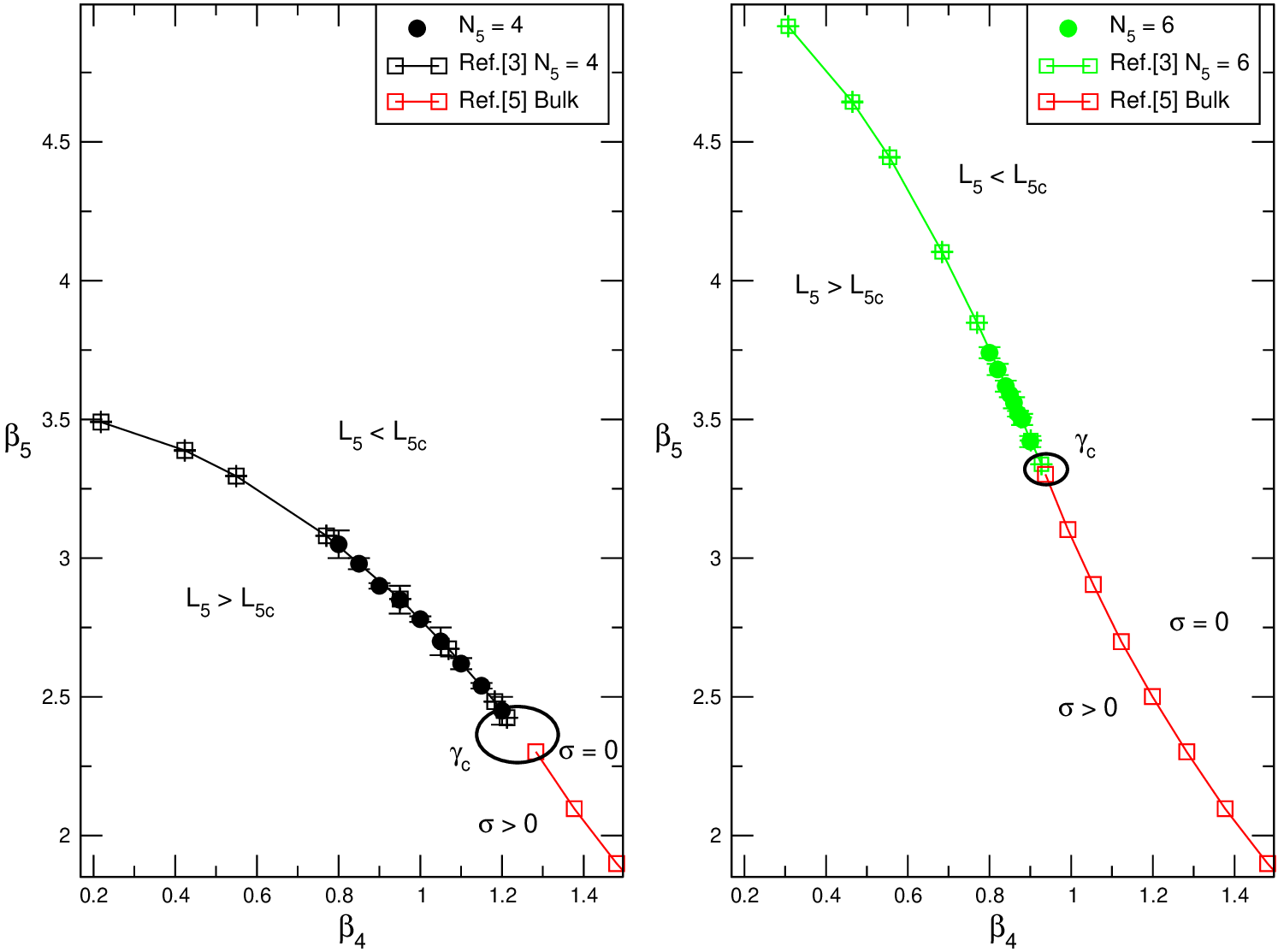}
 \caption{Phase diagram of the five dimensional SU($2$) pure gauge
   lattice model in the ($\beta_4,\beta_5$) plane for different values
   of $N_5$. The bulk phase transition separating a confinement from a
   Coulomb--like phase disappears for $\gamma > \gamma_c$ into a
   physical thermal--like phase transition. The location of this
   transition changes in the parameter space as we change $N_5$.}
 \label{fig:phase_diagram_splitview}
\end{figure}

\begin{figure}[t]
 \centering
  \begin{tabular}{cc}
    \includegraphics[width=0.47\textwidth]{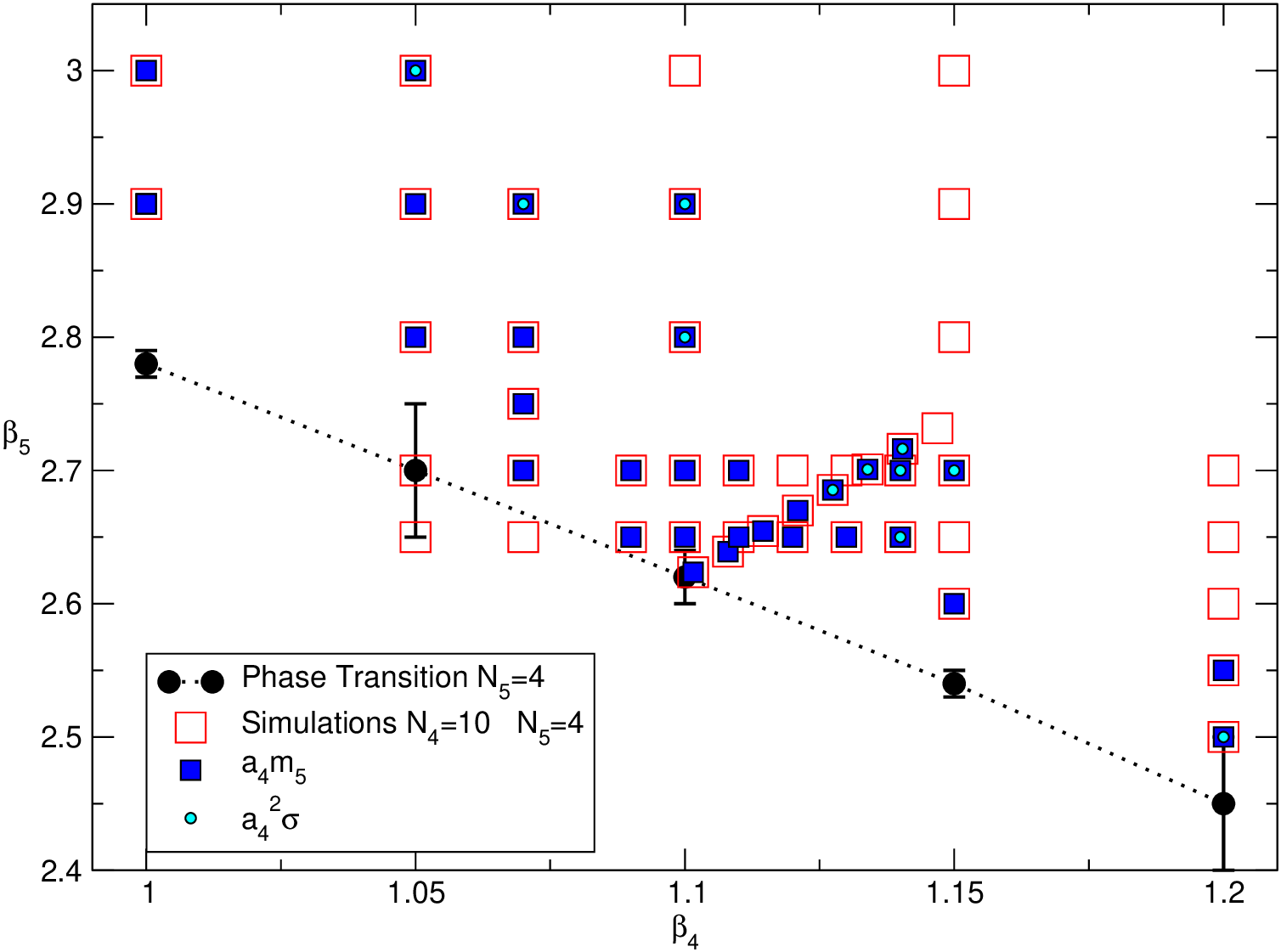}
    &
    \includegraphics[width=0.47\textwidth]{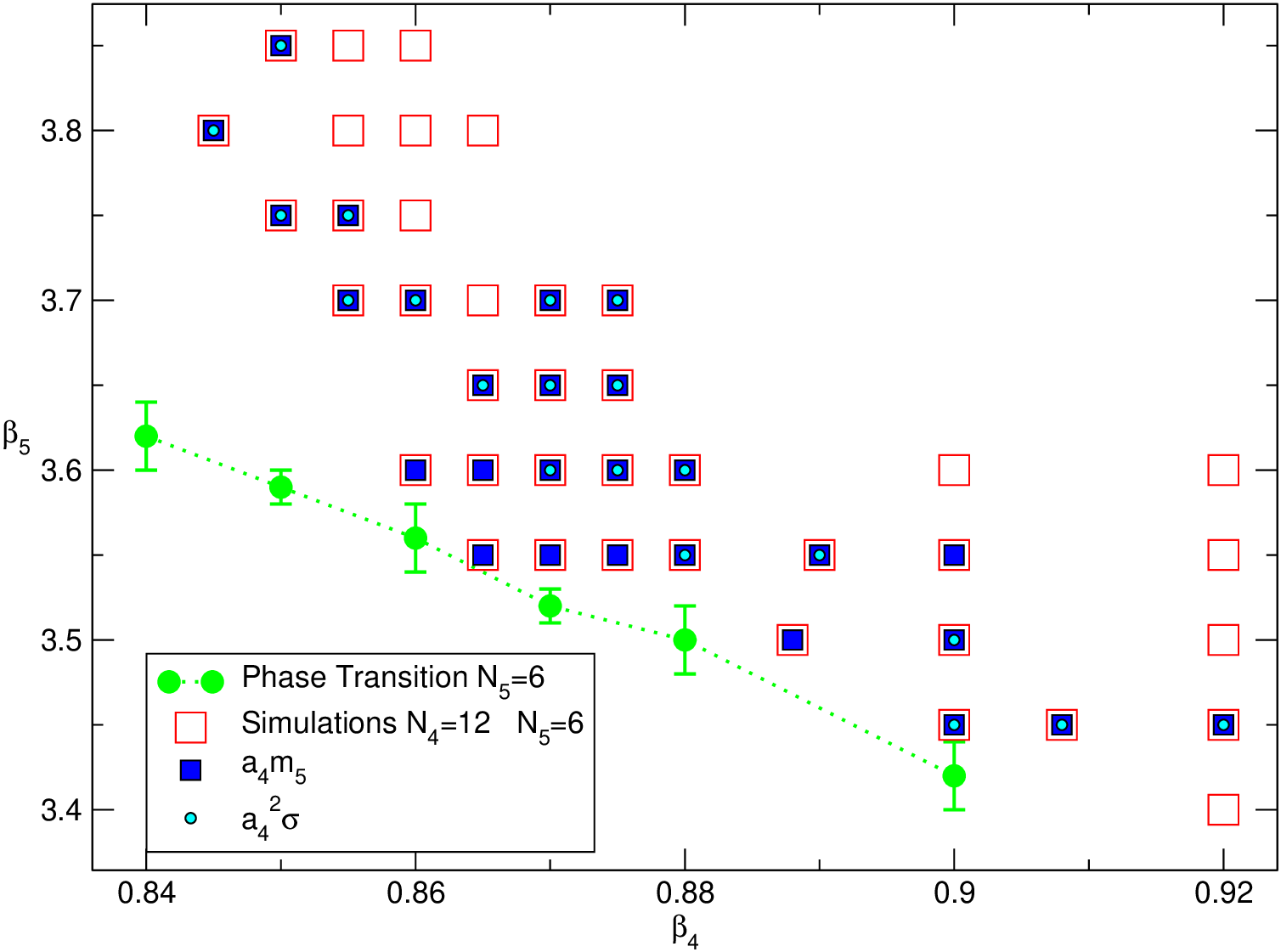}
    \\
  \end{tabular}
  \caption{The plots show the region of the phase diagram that we
    explored with 
    numerical simulations, both for $N_5=4$ (left) and $N_5=6$
    (right). The location of the second order phase transition is also
    shown.
    The blue squares are points where the scalar mass $a_4m_5$ was
    reliably extracted, whereas 
    the green circles represent points where we were able to measure
    the string tension $a_4\sqrt{\sigma}$.} 
  \label{fig:points}
\end{figure}

\begin{figure}[t]
 \centering
  \begin{tabular}{cc}
    \includegraphics[width=0.47\textwidth]{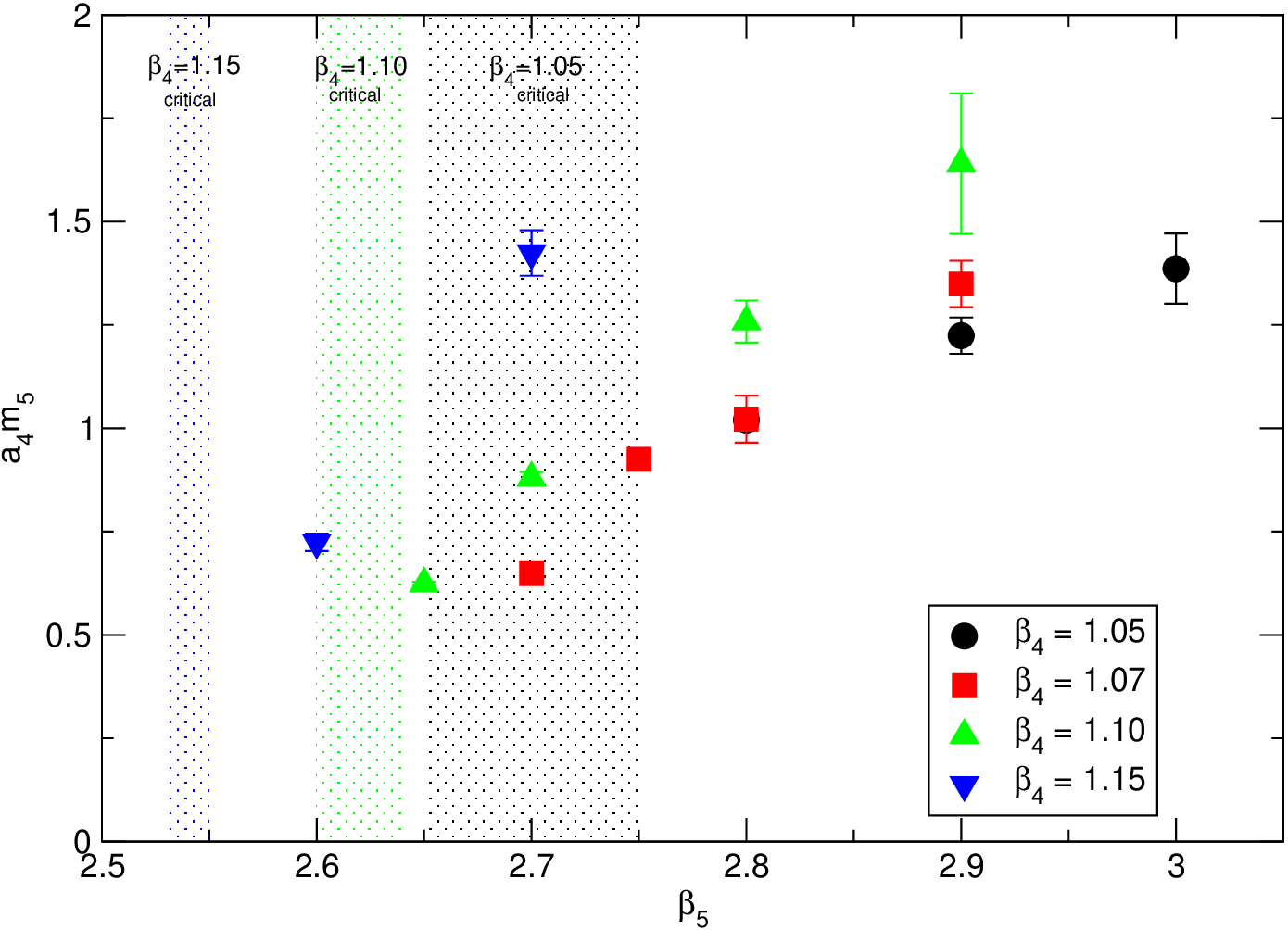}
    &
    \includegraphics[width=0.47\textwidth]{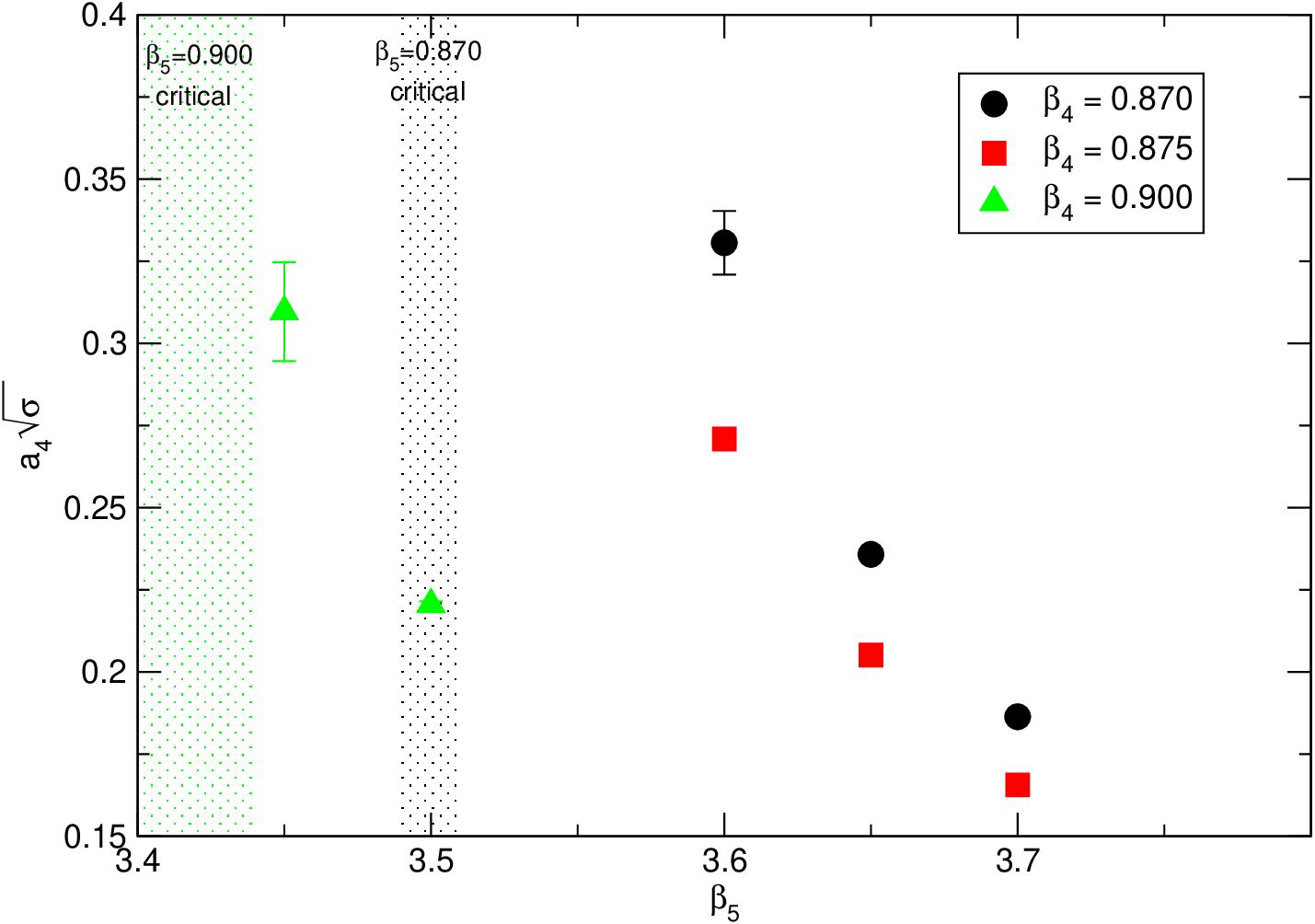}
    \\
  \end{tabular}
  \caption{(left) The scalar mass in units of the lattice spacing
    $a_4m_5$ as a function of $\beta_5$ and for four different values
    of $\beta_4$ at $N_5=4$. (right) The string tension in units of the lattice spacing
    $a_4\sqrt{\sigma}$ as a function of $\beta_5$ and for three different values
    of $\beta_4$ at $N_5=6$. In both panels the approximate location of the critical
    region is shown.} 
  \label{fig:scalar-string}
\end{figure}

\begin{figure}[b]
 \centering
  \begin{tabular}{cc}
    \includegraphics[width=0.47\textwidth]{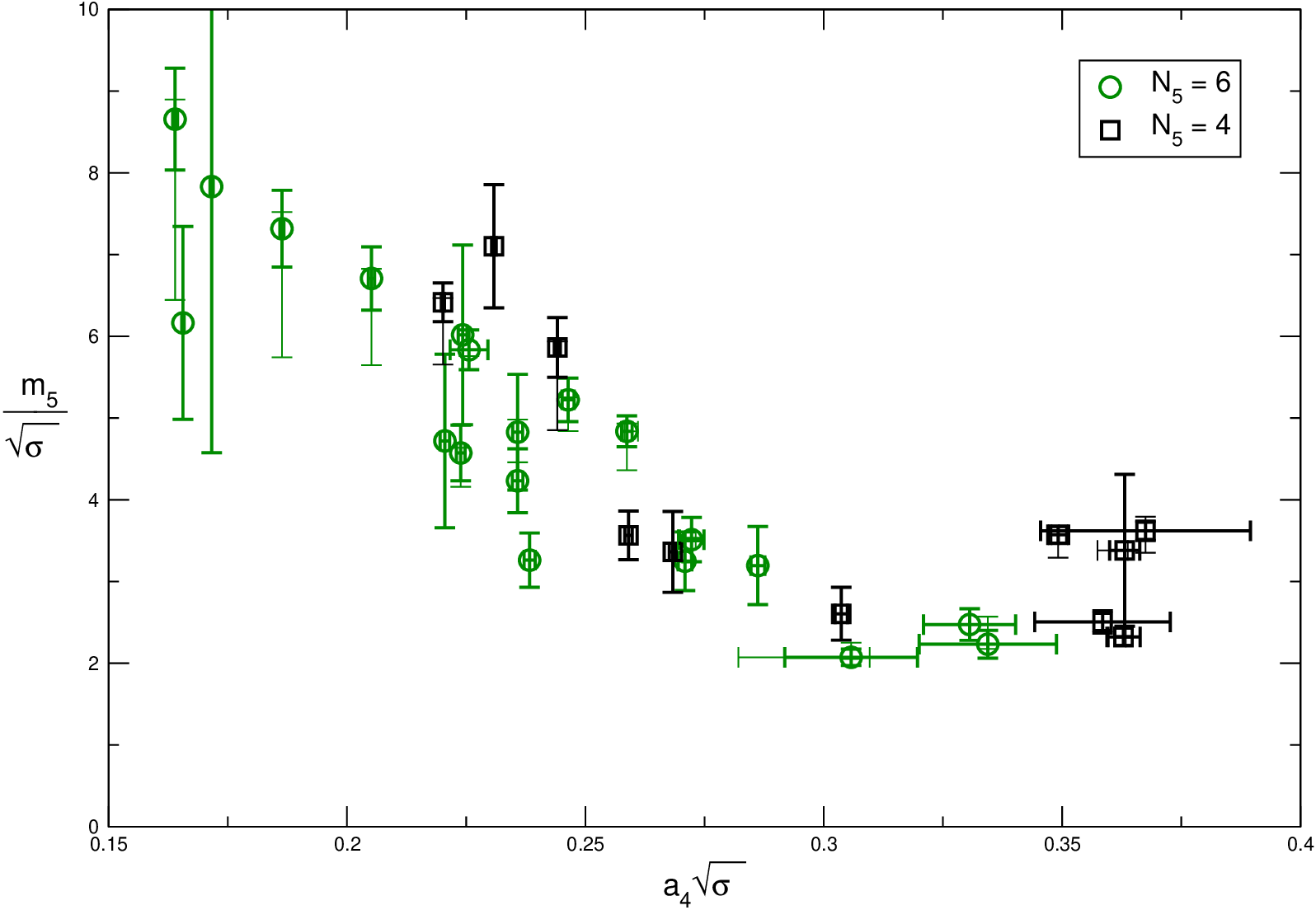}
    &
    \includegraphics[width=0.47\textwidth]{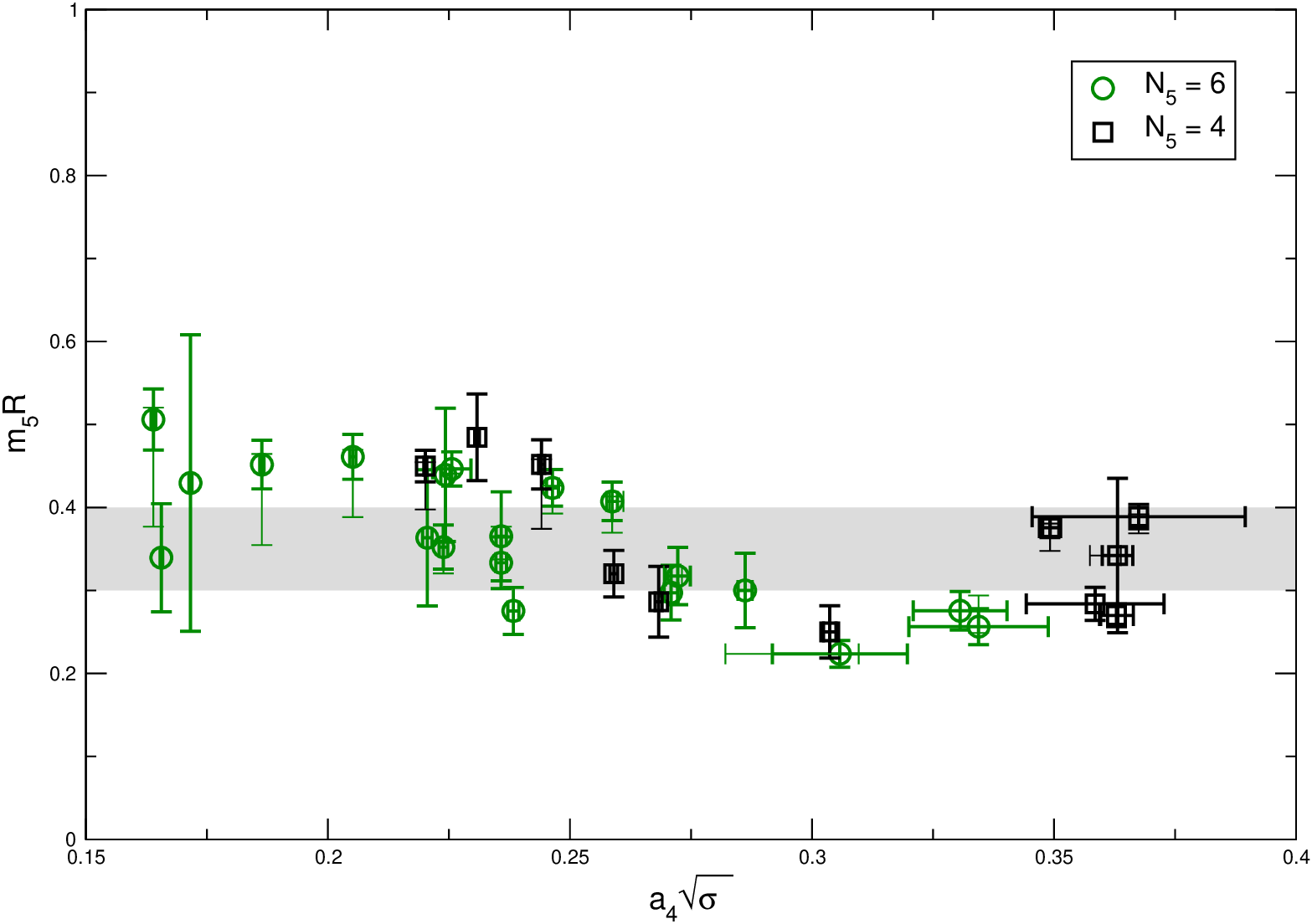}
    \\
  \end{tabular}
  \caption{In this plot we show the dependence on the cut--off length
    given by the lattice spacing $a_4 \sqrt{\sigma}$ for two
    observables: (left) the scalar mass in
    units on the string tension 
    and (right) the scalar mass in units of the
    compactification radius.
    The grey band includes all points within two standard
    deviations. When present, systematic 
    errors are shown with thinner error bars, whereas the thicker ones
    represent standard deviations.} 
  \label{fig:scalarR-string}
\end{figure}

Since it is crucial for our purposes to simulate the theory in the
correct phase, we briefly discuss the current understanding of the
phase diagram of the SU($2$) pure gauge theory in five dimensions
described by the action in
Eq.~\eqref{eq:aniso-lattice-action-E}. Results are available both at
$\gamma > 1$~\cite{ejiri,philippe} and $\gamma\leq
1$~\cite{antonio}.\\
The isotropic model, where the lattice spacings are the same $a_4 =
a_5$, has a bulk phase transition when all the dimensions are equal
and large. The bulk line separates a confined phase ($\sigma > 0$)
that is connected to the strong coupling regime from a Coulomb--like
phase ($\sigma = 0$) connected to the weak coupling. This bulk
transition disappears when the lattice size in anyone dimension is
decreased below a critical size, $L_{c}$, which is the critical length
of the Polyakov loop in that direction. Below $L_c$ center symmetry is
broken. In this case the phase transition becomes a second order one
in the same universality class of the four--dimensional Ising model. We
take $L_5$ to be our compactified length and in
Fig.~\ref{fig:phase_diagram_splitview} we show the pattern of phase
transition for $L_5=4a_5$ and $L_5=6a_5$ in the region $\gamma >
1$. The main feature is that at fixed lattice geometry the nature of
the phase transition line strongly depends on the anisotropy: the
second order phase transition related to centre breaking merges into
the bulk phase transition when
$\gamma \lesssim \gamma_c$. The emerging physical picture tells us
that the disappearance of the bulk phase transition happens as soon as
the five--dimensional system compactifies; in other words, $\gamma_c$
defines a critical lattice spacing in the extra dimension $a_{5c}$
that makes $2 \pi R = a_5 N_5$ smaller than the critical $L_{5c} =
a_{5c} N_5$.
The interesting region for our purposes, is at $\gamma >
\gamma_c$ and above the line of second order phase transition, where
the extra dimension is smaller than its critical value $L_5 <
L_{5c}$. We refer to this phase as the \emph{dimensionally reduced}
phase, following Ref.~\cite{philippe}.\\
Since this is the first time that this particular region of the phase
space is explored with lattice simulations, we performed a broad scan,
aiming primarily at identifying the interesting region. As a
consequence, there are lattices for which we were unable to measure
precisely both the string tension $a_4\sqrt{\sigma}$ and the scalar
mass $a_4m_5$. In Fig.~\ref{fig:points}, we show all the simulations'
points and at the same time we identify the ones where either $a_4m_5$
or $a_4\sqrt{\sigma}$ could not be extracted satisfactorily from
correlation functions of suitable gauge invariant operators.\\
The emergent pattern seems to suggest that the lattice spacing $a_4$
changes dramatically in these regions of the phase space.  For
example, looking at Fig.~\ref{fig:points}, we notice that the string
tension $a_4\sqrt{\sigma}$ can only be measured in a small subset of
points; the points closer to the line of second order phase transition
are characterised by spatial Polyakov loops whose mass is too high for
a signal to be extracted reliably. Since the mass of the loops is
given by $N_4 a_4 \sigma$, in this region the lattice
spacing $a_4$ is getting larger in units of the string tension.
Following our discussion in Sec.~\ref{sec:lattice-setup}, we
regard the region close to the line of phase transition as the one
characterised by a small cut--off $\LambdaUV$. In this region, there
is not a clear separation between the low--energy physics and the
cut--off, and we expect to observe large discretisation errors. To
make things even more interesting, we find the scalar mass $a_4m_5$ to
be small in this same region, where $a_4$ is large. In fact, it
turns out to be very difficult to find points in the phase diagram
where both $\sqrt{\sigma}$ and $m_5$ are separated from the cut--off
scale at the same time. This results in a  scalar mass $m_5 \sim
\sqrt{\sigma}$ for all the points on which we were able to reliably
measure the string tension. On the other hand, a \emph{light} scalar
does exist very close to the second order transition line,
where $a_4m_5$ is small and $a_4\sqrt{\sigma}$ is large.\\
A more quantitative statement can be made by looking at the measured
observables as functions of the bare parameters. For example, our data
allow us to study the behaviour of $a_4\sqrt{\sigma}$ at fixed value
of $\beta_4$ as we change $\beta_5$, and viceversa. The same can be
done with $a_4m_5$. In the left panel of Fig.~\ref{fig:scalar-string} we see
that the scalar mass approaches the cut--off scale $a_4m_5
\gtrsim 1$ as we move away from the line of second order phase
transition at fixed $\beta_4$. While the scalar mass becomes smaller
as we approach the critical line, the opposite happens to the string
tension, as it is shown in the right panel of
Fig.~\ref{fig:scalar-string}.
Each point we have simulated on the phase
diagram correspond to a precise location in the space given by the
three energy scales $\LambdaUV$,
$\LambdaR$ and $m_5$. We can therefore translate our results at
$N_5=4$ and $N_5=6$ into a common set of points
$(\LambdaUV,\LambdaR,m_5)$. This approach allows us to study $m_5$ as
a function of the other two energy scales, instead of the bare
parameters. From now on we express the energies $\LambdaUV$ and
$\LambdaR$ using their length counterpart, $a_4\sqrt{\sigma}$ and
$R\sqrt{\sigma}$ respectively. This two length scales are related to
each other by Eq.~\eqref{eq:separation-lattice} and they are both
measured non--perturbatively: the first is directly measured, whereas
the second relies on our interpolation of $\xi$ from
Ref.~\cite{ejiri}. With our available data, we can explore the
behaviour of the scalar
mass $m_5$ in the following region of lattice spacings $a_4$ and
compactification radii $R$ 
\begin{equation}
  \label{eq:scale-interval}
  0.15 \; < \; a_4\sqrt{\sigma} \; < \; 0.40
  \; , \qquad
  0.05 \; < \; R\sqrt{\sigma} \; < \; 0.12
  \point
\end{equation}
The major advantage of interpreting the data in this new physical
space is that we can disentangle compactification effects from
cut--off effects. Since our values for the lattice spacing usually
correspond to
different compactification radii, we look at the combination $m_5R$. 
This is expected to be independent on $R$, while
retaining any dependence on the cut--off. If
Eq.~\eqref{eq:one-loop-mass} holds, then $m_5R$ should be
independent of $R$ and $a_4$ at leading order. In
Fig.~\ref{fig:scalarR-string} we plot $m_5/\sqrt{\sigma}$ as a
function of
$a_4\sqrt{\sigma}$ in the left panel, and $m_5R$ in the right
panel. The observed range for
the scalar mass in units of the string tension is
\begin{equation}
  \label{eq:m5sigma-interval}
  2 \; < \; m_5/\sqrt{\sigma} \; < \; 10
  \comma
\end{equation}
whereas the scalar mass in units of the
compactification radius is in the range
\begin{equation}
  \label{eq:m5R-interval}
  0.2 \; < \; m_5R \; < \; 0.5
  \point
\end{equation}
The latter range is smaller by
almost one order of magnitude for same interval of lattice
spacings. This evidence support the observation that the dependence on
$a_4\sqrt{\sigma}$ is mild.

\section{Conclusions}
\label{sect:conclusions}
In this work we presented a non--perturbative study of pure SU($2$)
gauge theory in five dimensions. If the scales of the theory are
properly separated, we expect the 
low-energy dynamics of this theory to describe a four-dimensional
gauge theory coupled to a scalar field. We have measured the mass of
the scalar particle in a specific region of the bare parameters space,
where we expect to find the desired separation between physical
scales. We have also determined numerically the four--dimensional
lattice spacing in units of the string tension.\\
The final picture seems to confirm the
possibility of effectively describing a four--dimensional Yang--Mills
theory with a scalar adjoint particle in the continuum limit.
Even though the search for a light scalar requires fine tuning in this
simple model, we have shown that its mass is only very mildly affected
by the ultra--violet cut--off, whereas it strongly depends on the
radius of the compactified extra dimension. This is entirely
compatible with the perturbative result of
Eq.~\eqref{eq:one-loop-mass} and it is the first non--perturbative
evidence that the mass of scalar particles coming from a
compactification mechanism does not have a quadratic dependence on the
cut--off.

%-----------------------BIBLIOGRAPHY----------------------------

\end{document}